\documentstyle[epsf,psfig,twocolumn,aps,prl,floats]{revtex}
\begin{document}

\draft
\title{Numerical Study of the Localization-Delocalization Transition
for Vibrations in Amorphous Silicon}

\author{ William Garber and Folkert M. Tangerman}
\address{Department of Applied Math and Statistics, 
State University of New York, Stony Brook, New York 11794-3600}

\author{ Philip B. Allen}
\address{Department of Physics and Astronomy, State University of New York,
Stony Brook, New York 11794-3800}
 
\author{Joseph L. Feldman}
\address{Naval Research Laboratory, Washington DC 20375-5345}

\maketitle

\begin{abstract}

Numerical studies of amorphous silicon in harmonic approximation show that the
highest 3.5\% of vibrational normal modes are localized.
As vibrational frequency increases through the boundary
separating localized from delocalized modes, near $\omega_c$=70meV,
(the ``mobility edge'')
there is a localization-delocalization (LD) transition,
similar to a second-order thermodynamic phase transition.
By a numerical study on a system with 4096 atoms,
we are able to see exponential decay lengths of exact vibrational
eigenstates, and test whether or not these diverge at $\omega_c$.
Results are consistent with a localization length $\xi$
which diverges above $\omega_c$ as $(\omega-\omega_c)^{-p}$ where
the exponent is $p \sim 1.3 \pm 0.5$.
Below the mobility edge we find no evidence for a diverging
correlation length.  Such an asymmetry would contradict scaling ideas,
and we suppose it is a finite-size artifact.  If
the scaling regime is narrower than our ($\approx$ 1 meV)
resolution, then it cannot be seen directly on our finite system.

\end{abstract}


Harmonic normal modes of vibration can be classified
as extended or localized.  In dimension $d=3$, the vibrational
spectrum has sharp boundaries (``mobility edges'')
separating these two kinds of modes.  The sudden change
at the mobility edge is a ``localized to delocalized''
(LD) transition (Anderson, 1958).  Anderson's
paper considered localization of electronic wavefunctions,
and most subsequent work focussed on this case
(Lee and Ramakrishnan, 1985;
Kramer and MacKinnon, 1993;
Belitz and Kirkpatrick, 1994; Ohtsuki, Slevin, and Kawarabayashi, 1999).
Harmonic vibrations (treated classically --
quantization does not matter for localization)
have similar wave properties to electrons in independent-electron
approximation.  However, in one respect, 
vibrational states differ from electron states.
Long wavelength, low frequency (sound) modes always exist, extending
throughout a homogeneous sample, and are thus delocalized.
There is no analog of this ``gapless'' mode for electrons.
Nevertheless, it is generally believed that at higher frequencies, vibrations
and electrons have similar localization properties
(Kirkpatrick, 1985; Sheng, Zhou, and Zhang, 1994; Sheng, 1995;
Akita and Ohtsuki, 1998).

In glasses, only a few percent of normal modes can propagate ballistically
like sound.  As frequency increases, ballistic character is lost 
at typical frequencies of 10 meV (Rat {\sl et al.} 1999), but the LD transition
(which is hard to probe by direct experiment) is postponed to a much
higher frequency $\omega_c$ (Allen and Feldman 1989; Feldman {\sl et al.} 1991; 
Sheng, Zhou, and Zhang 1994; Finkemeier and von Niessen 1998;
Pilla {\sl et al.} 2000.) 
Recent preprints by Galanti and Olami (1999) and Gat and Olami (2000)
have examined the nature of vibrational states near the mobility edge.
Using scalar vibrations, they
find numerical evidence that vibrations may belong to
a different ``universality class'' from electrons.

The concerns of the present paper
are not so much the formal questions of universality,
but instead specific questions of how vibrational eigenstates
look in real materials.  Experiment does not give access 
to vibrational eigenvectors or to electron wavefunctions.
Much of the information about these states comes from computer
studies.  Most of these studies use simplified ``toy'' models,
such as ``scalar vibrations'' where the vector displacement
of an atom is replaced by a scalar variable, or hypothetical
structures with fractal geometries (Alexander and Orbach, 1982;
Nakayama {\sl et al.}, 1994; Kantelhardt {\sl et al.}, 1998).  
Here we look at properties of vibrational eigenstates near the mobility
edge in the most realistic model presently available for amorphous Si.
We use vector vibrations and find
that the localization length diverges in the usual way as
the mobility edge is approached from the localized side.
We find no evidence for a diverging correlation length
on the delocalized side.  These results seem to agree with
those of Gat and Olami but not with those of Galanti and Olami.
We believe that our results are reliable, 
accurately portraying how vibrations look in our model, and perhaps
generally in glasses, at energies distant by more than 1 meV from
the mobility edge.  It is likely that for energies distant by
less than 1 meV from the mobility edge, correlation lengths
diverging with a universal critical exponent occur on both sides
of the mobility edge, but this information is outside our grasp.

The usual method for extracting information about the critical
regime from finite systems in $d=3$ is to use
finite-size scaling combined with calculations averaged over
large ensembles (Kramer and MacKinnon 1993).
By this method one is able indirectly to extract information
from states whose localization length is comparable to the
sample size, provided that the scaling hypothesis is accepted.
Our method is completely different.  Our system is large
($L=44\AA$) and has more than 400 eigenstates which are well-localized
within this system size.  We can look directly at these eigenstates
to see whether they decay exponentially.  Averaging over groups
of states in an energy interval of 1 meV enhances the quality
of the measured decay constants, but even single states can
be examined and measured.  For the case of glasses with only
a few percent of states localized, the mobility edge is
easy to locate (to within 1 meV) without ensemble averaging.
This was also found by Sheng, Zhou, and Zhang (1994).
The direct proceedure is complementary 
to, and provides a necessary reality check on the usual
scaling methods.  It does not work well for
simpler model systems or systems with very large disorder.
For example, Feldman {\sl et al.} (1993) artificially
pushed $\omega_c$ deeper into the spectrum of amorphous Si by 
adding random mass disorder.  At energies a few meV distant from
the mobility edge, eigenstates were not sharply localized within the
finite sample, and the clarity of location of $\omega_c$ was lost.

We study amorphous silicon in harmonic approximation,
using a realistic model which we have already studied
extensively.  Many of these studies are summarized
in a paper hereafter referred to 
as I (Allen {\sl et al.} 1999).  
Amorphous Si has approximate tetrahedral coordination,
and can be called an ``over-constrained network glass.''
Even though it is not an especially typical glass, we
suspect the vibrational properties under discussion here
have much in common with other glasses.
Atomic coordinates were generated by Wooten
using the algorithm of Wooten, Weiner, and Weaire (1985),
with 4096 atoms in a cubic box of side $L=44\AA$, 
continued periodically to infinity.  
The coordinates were relaxed
to a local minimum of the Stillinger-Weber potential (Stillinger and
Weber, 1985), which is then used to find the harmonic restoring
forces.  Eigenvectors and
eigenfrequencies were computed on the Linux cluster Galaxy 
( ref: http://www.galaxy.ams.sunysb.edu) using the Scalapack (ref:
http://www.netlib.org/scalapack)
routine PDSYEV.  The calculation used
20 Pentium II (400 Mhz) processors and required CPU time of 4.5 hours
(including verification of the correctness of the result).
The distribution of eigenfrequencies
(which are all real) is shown in Fig. 1 of I.  
Finkemeier and von Niessen (1998) have made a systematic
study of this model and several related ones, including more
detailed size-dependence than we have done, as well as some
variation with disorder.  Their results are for vibrational
densities of states and localization properties are very
similar to ours where they overlap.  In addition, they discovered
a surprising insensitivity of the location of the mobility edge
to the degree of amorphous disorder introduced in the model.

The definition of a localized state is exponential decay of
the eigenvector with distance from some center $\vec{R}_0$:
\begin{equation}
|\vec{\epsilon}_i(\vec{R}_n)|^2
=|a_i(\vec{R}_n)|^2\exp(-2|\vec{R}_n -\vec{R}_0|/\xi_i)
\label{eq:xi}
\end{equation}
where $a_i(\vec{R}_n)$ isolates a factor which
is not systematic but instead fluctuating in magnitude and sign.
These fluctuations are random in the sense that one expects an
ensemble of similarly prepared systems to show eigenstates for which
the localization length $\xi_i$ has ``universal'' behavior near the mobility
edge, and the deviations $a_i-\overline{a}$ from exponential fall-off to be 
``pseudo-random'' with zero mean ($\overline{a}=0$) and distribution
$P(a)$ independent of $\vec{R}_n$.
Representative plots are shown in Fig. 5 of I.
In order to confirm numerically that exponential decay is occuring, 
it is desirable to have $|\vec{\epsilon}_i|^2$ decrease by at least 100
over the maximum available $|\vec{R}_n -\vec{R}_0|$ which is $L/2$
or 22$\AA$.  This means that accurately determined
values of $\xi$ are 8$\AA$ or 
less.  Very large system sizes are needed to measure the large
values of $\xi$ expected close to the mobility edge.

\par
\begin{figure}[t]
\centerline{\psfig{figure=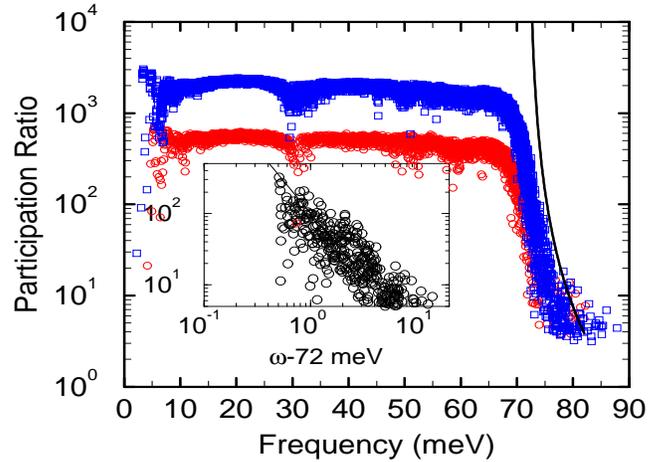,height=2.5in,width=3.6in,angle=0}}
\caption{Participation ratio for normal modes
of the 4096-atom model (higher lying data, plotted as squares)
and for the 1000-atom model (lower lying data, plotted as circles)
plotted versus mode energy.  The thick solid line seen at higher energy
is the scaling result using the dashed line from Fig. 2
as discussed in the text.  The inset shows the higher frequency data
for the 4096-atom model on a logarithmic frequency scale, and a straight
line whose equation is $p \propto (\omega - 72 {\rm meV})^{-3p}$ with
exponent $p=0.6$.}
\label{fig:partrat}
\end{figure}

Our previous studies on various size models have shown that most
properties are reproducible from model to model.  Figure 1
shows how reproducible is the participation ratio $p_i$ defined
as (Bell and Dean, 1970)
\begin{equation}
p_i = \frac{[\sum_n |\vec{\epsilon}_i(\vec{R}_n)|^2]^2  }
    {\sum_n |\vec{\epsilon}_i(\vec{R}_n)|^4}.
\label{eq:pr}
\end{equation}
The numerator is simply the normalization factor, equalling 1 for
normalized vectors.  If the mode $i$ is a pure plane wave,
the denominator sums to $1/N$, so $p_i$ is the number of atoms, $N$. 
If the mode $i$ is localized on a single site,
the denominator is 1 and $p_i=1$.  Roughly speaking,
$p_i$ measures the number of atoms which participate in the vibration.
To be more quantitative, the normalization integral can
be written as 
\begin{equation}
1=\int d\vec{r}\rho(\vec{r})|a_i(\vec{r})|^2 e^{-2r/\xi} 
\label{eq:norm}
\end{equation}
where the integral goes over the volume $V$ of a 4096-atom cell,
and the density $\rho(\vec{r})$ is $\sum_n \delta(\vec{r}-\vec{R}_n)$.
Approximating the density as $N/V$ and $|a_i(\vec{r})|^2$ by its
average $\bar{a^2}$, for a state decaying well within one
cell volume we obtain $\overline{a^2} \approx (L/\xi)^3/\pi N$.  
Making the same
approximations to evaluate the denominator of Eq.(\ref{eq:pr}),
we obtain
\begin{equation}
p \approx 8\pi N(\xi/L)^3 \overline{a^2}^2/\overline{a^4} 
   \approx (8\pi N/3)(\xi/L)^3,
\label{eq:prsc}
\end{equation}
where the assumption is made that the random numbers $a_i$ have a Gaussian
distribution.

According to scaling theory, eigenvectors for states near
the mobility edge (which serves as the critical point) should
be characterized by a correlation length.  On the 
localized side of the mobility
edge, the decay length serves as the correlation length.  
The correlation length should diverge
upon approaching the mobility edge.  In the ``critical region''
the divergence is characterized by an exponent $\nu$,
\begin{equation}
\xi \propto |\omega-\omega_c|^{-\nu} \Rightarrow p
  \propto |\omega-\omega_c|^{-3\nu}.
\label{eq:expo}
\end{equation}
The exponent $\nu$ is supposed to be ``universal,'' that is,
the same for all systems with time-reversal symmetry,
and is found to have the value $\nu = 1.57 \pm 0.02$ by numerical
study of electron models in three dimensions (Ohtsuki {\sl et al}, 1999).
The data for $p_i$ are sufficiently dense that we can test this.
We choose only modes with $\omega_i >$72.5meV, with $p_i <$300.
Making log-log plots of $p$ versus
$\omega-\omega_c$ for various choices of $\omega_c$
(see the inset to Fig. 1), the data
are roughly consistent with power laws, but precision is low and
it is not possible even to decide where the mobility edge lies with
much accuracy, much less to find a
critical exponent.  The data fit roughly to
$\xi\propto(\omega-69{\rm meV})^{-1.5}$,  $(\omega-70{\rm meV})^{-1.2}$, 
$(\omega-71{\rm meV})^{-0.75}$, and $(\omega-72{\rm meV})^{-0.6}$.
Therefore we try a different method which uses more of the information
in the eigenvectors.

For each mode in the spectrum we have
located the atom $\vec{R}_0$ on which $|\vec{\epsilon}_i|^2$ is maximum,
and computed the correlation function
\begin{equation}
f_i(R)=\sum_n |\vec{\epsilon}_i(\vec{R}_n)|^2 \overline{\delta}
(R-|\vec{R}_n -\vec{R}_0|)/
\sum_n \overline{\delta}(R-|\vec{R}_n -\vec{R}_0|)
\label{eq:eigfall}
\end{equation}
where $\overline{\delta}(x)$ is a rectangularly shaped approximation
to a delta function with width 0.2$\AA$.  The correlation functions
$f_i(R)$ were then averaged over all states $i$ with $\omega_i$
in intervals ($\omega,\omega+d\omega$) of width $d\omega$=1 meV.
This means typically 100 states were averaged.
These averaged correlation functions $f(\omega,R)$ were then plotted
semilogarithmically over the spatial range $0<R<20\AA$.  For
all $\omega>72$ meV, the semilog graph fell linearly (with random
fluctuations) over a range $R_1<R<20\AA$, and decay constants
$\alpha$ and errors $\Delta\alpha$ could be found.
The distance $R_1$ below which non-exponential
behavior occurred was about 13$\AA$ for the frequency range
72-73 meV, just above $\omega_c$, and decreased steadily to
4$\AA$ as $\omega$ increased.
Values of $\alpha$ and errors are shown in Fig. 2.

\par
\begin{figure}[t]
\centerline{\psfig{figure=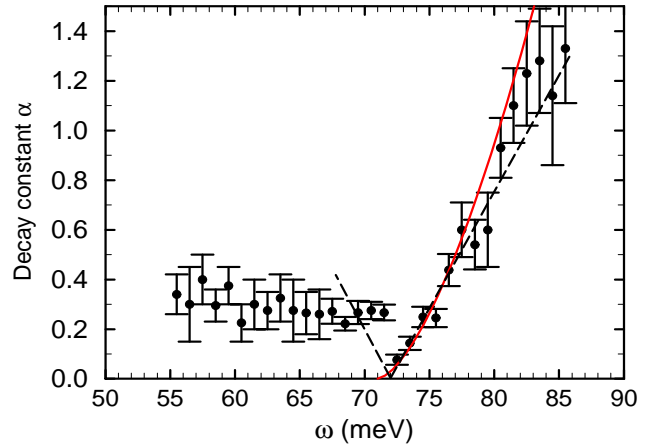,height=3.0in,width=3.6in,angle=0}}
\caption{Decay constant $\alpha=2/\xi$ in reciprocal Angstroms
{\sl versus} energy $\omega$ in meV 
 for vibrations in amorphous silicon.  Dashed lines show scaling behavior 
 with exponent $\nu$=1.  The curved solid line shows scaling
 behavior with $\nu$=1.57}
\label{fig:decay}
\end{figure}

The data for $\omega>$72meV are consistent with a linear relation
$2/\xi=0.094(\AA{\rm \ meV})^{-1}(\omega-72{\rm \ meV})$,
that is, with exponent $\nu=1$.  These data are also consistent
with the curve $2/\xi=0.03(\AA{\rm \ meV}^{1.57})^{-1}
(\omega-71{\rm \ meV})^{1.57}$,
that is, with exponent $\nu=1.57$.  
These results can be used in Eq.(4)
to model the behavior of $p_i$ for $\omega_i > 72$meV.  The linear
fit is shown as the solid line on Fig. 1.  A much improved fit 
to the data in Fig. 1 is obtained
by shifting the mobility edge down from 72meV to 70meV.  The disagreement
is surprisingly large.  Using the fit with $\nu=1.57$ provides no improvement.
The explanation is probably that $p_i$ is dominated by exponential decay
(as assumed in Eq. (\ref{eq:prsc})) only very near $\omega_c$ where larger
system sizes are needed.
This problem is avoided when we directly examine 
the spatial behavior of the eigenstates.

The correlation length
for delocalized states gives the length scale over which fluctuations
in magnitude of the eigenvector decay back to the average value,
\begin{equation}
|\vec{\epsilon}_i(\vec{R}_n)|^2 = 
|a_i(\vec{R}_n)\exp(-|\vec{R}_n -\vec{R}_0|/\xi_i)
+b_i(\vec{R}_n)|^2
\label{eq:deloc}
\end{equation}
where $a_i(\vec{R}_n)$ and $b_i(\vec{R}_n)$ are both pseudorandom
with zero mean.  For a normalized state, $\overline{b_i^2}$ is 1/N.
The same correlation function (Eq.(6)) was computed and averaged
for states in the range 55--72 meV.  The results agree qualitatively
with the form 
\begin{equation}
\overline{f(R)}=\overline{a^2}\exp(-2R/\xi)+1/N
\label{eq:fbar}
\end{equation}
which is expected if $a_i$ and $b_i$ are uncorrelated.  The decay
constants $\alpha=2/\xi$ are plotted on Fig. 2.  The value of $\xi$
seems fixed near 7$\AA$.  This result contradicts the
expectation of scaling theory that $\xi$ should diverge with the
same exponent on both sides of the mobility edge.  

We believe that the numerical data of Fig. 2 represent reliable
values of exponential decay lengths for our model of amorphous
Si.  In particular, these data show clearly that the
mobility edge coincides with a point of apparent divergence
of this length on the localized side, and also clearly show an
asymmetry between the localized and delocalized side of $\omega_c$.
Unfortunately, there is no way from this study to decide whether
the measured lengths are close enough to the critical region to
test the applicability of scaling theory to this system.
Nevertheless, the surprising asymmetry seems to be a real property, similar
to what was seen by Galanti and Olami (1999).

\acknowledgements

We thank J. Fabian and B. Nikolic for guidance.  We thank F.
Wooten for building the model, and S. Bickham and
K. Soderquist for relaxing the coordinates.
WG acknowledges support from USAFOSR (F496209510387 and F496209510407);
FT acknowledges support from DOE (DEFG0398DP00206);
PBA was supported in part by NSF grant no. DMR-9725037;
JF was supported by the U. S. Office of Naval Research.

\begin{center}
{\bf REFERENCES}
\end{center}

\noindent		Akita, Y. and Ohtsuki, T., 1998,
			J. Phys. Soc. Jpn. {\bf 67}, 2954.\\

\noindent		Alexander, S., and Orbach, R., 1982,
			J. Phys. Lett. (Paris) {\bf 43}, L625.\\

\noindent		Allen, P. B. and Feldman, J. L., 1989,
        		Phys. Rev. Lett. {\bf 62}, 645.\\

\noindent		Allen, P. B., Feldman, J. L., Fabian, J.,
			and Wooten, F., 1999,
			Phil. Mag B {\bf 79}, 1715.\\

\noindent	        Anderson, P. W., 1958,
                        Phys. Rev. {\bf 109}, 1492.\\

\noindent		Belitz, D. and Kirkpatrick, T. R., 1994,
			Rev. Mod. Phys. {\bf 66}, 261.\\

\noindent		Feldman, J. L., Kluge, M. D., Allen, P. B.,
			and Wooten, F., 1993,
        		Phys. Rev. B {\bf 48}, 12589.\\

\noindent		Finkemeier, F. and von Niessen, W., 1998,
			Phys. Rev. B {\bf 58}, 4473.\\

\noindent		Galanti, B. and Olami, Z., 1999,
			cond-mat/9909227.\\

\noindent		Gat, O. and Olami, Z., 2000,
			Phys. Rev. E (in press).
			cond-mat/9911409.\\

\noindent		Kantelhardt, J. W., Bunde, A., and Schweitzer, L., 1998,
			Phys. Rev. Lett. {\bf 81}, 4907.\\

\noindent		Kirkpatrick, T. R., 1985,
			Phys. Rev. B {\bf 31}, 5746.\\

\noindent		Kramer, B. and MacKinnon, A., 1993,
			Rep. Prog. Phys. {\bf 56}, 1469.\\

\noindent		Lee, P. A., and Ramakrishnan, T. V., 1985,
			Rev. Mod. Phys. {\bf 57}, 287.\\

\noindent		Nakayama, T., Yakubo, K., and Orbach, R. L., 1994,
			Rev. Mod. Phys. {\bf 66}, 381.\\

\noindent		Ohtsuki, T., Slevin, K., and Kawarabayashi, T., 1999,
			Ann. Phys. (Leipzig) {\bf 8}, 655.\\

\noindent		Pilla, O., Cunsolo, A., Fontana, A., Masciovecchio, C., 
			Monaco, G., Montagna, M., Ruocco, G., 
			and Scopigno, T., 2000,
			Phys. Rev. Letters {\bf 85}, 2136.\\

\noindent		Rat, E., Foret, M., Courtens, E., Vacher, R.,
			and Arai, M., 1999,
			Phys. Rev. Letters {\bf 83}, 1355.\\

\noindent		Sheng, P., Zhou, M., and Zhang, Z.-Q., 1994,
			Phys. Rev. Letters {\bf 72}, 234.\\

\noindent	        Sheng, P., 1995,
                        {\sl Introduction to Wave Scattering, Localization,
                        and Mesoscopic Phenomena,}
                        (Academic Press, San Diego).\\

\noindent		Song, P. H., and Kim, D. S., 1996,
			Phys. Rev. B {\bf 54}, 2288.\\

\noindent               Stillinger, F. H., and Weber, T. A., 1985,
                        Phys. Rev. B {\bf 31}, 5262.\\

\noindent               Wooten, F., Winer, K., and Weaire, D., 1985,
                        Phys. Rev. Letters {\bf 54}, 1392.\\

\end{document}